\documentclass{aa}
\usepackage[varg]{txfonts}
\usepackage{hyperref}
\usepackage{graphicx}
\usepackage{supertabular}
\usepackage{enumitem}

\begin{document}

\title{The Spatial Evolution of Young Massive Clusters}
\subtitle{III. Effect of the Gaia filter on 2D spatial distribution studies}

\author{Anne S.M. Buckner\inst{1}, Zeinab Khorrami\inst{2}, Marta Gonz\'alez\inst{3}, Stuart L. Lumsden\inst{4}, Paul Clark \inst{2} and Estelle Moraux\inst{5}}

\institute{School of Physics and Astronomy, University of Exeter, Stocker Road, Exeter EX4 4QL, UK \\ \email{a.buckner@exeter.ac.uk}
\and School of Physics and Astronomy, Cardiff University, The Parade, CF24 3AA, U.K. \and Universidad Internacional de Valencia (VIU),
C/Pintor Sorolla 21, E-46002 Valencia, Spain \and School of Physics and Astronomy, University of Leeds, Leeds LS2 9JT, U.K. \and  Universit\'e Grenoble Alpes, CNRS, IPAG, 38000 Grenoble, France}

\date{Received / Accepted}

\abstract{With the third release of the high-precision optical-wavelength Gaia survey, we are in a better position than ever before to study young clusters. However, Gaia is limited in the optical down to $G\sim21$\,mag, and therefore it is essential to understand the biases introduced by a magnitude-limited sample on spatial distribution studies.}{We ascertain how sample incompleteness in Gaia observations of young clusters affects the local spatial analysis tool INDICATE and subsequently the perceived spatial properties of these clusters.}{We created a mock Gaia cluster catalogue from a synthetic dataset using the observation generating tool MYOSOTIS. The effect of cluster distance, uniform and variable extinction, binary fraction, population masking by the point spread function wings of high-mass members, and contrast sensitivity limits on the trends identified by INDICATE are explored. A comparison of the typical index values derived by INDICATE for members of the synthetic dataset and their corresponding mock Gaia catalogue observations is made to identify any significant changes.}{We typically find only small variations in the pre- and post-observation index values of cluster populations, which can increase as a function of incompleteness percentage and binarity. No significant strengthening or false signatures of stellar concentrations are found, but real signatures may be diluted. Conclusions drawn about the spatial behaviour of Gaia-observed cluster populations that are, and are not, associated with their natal nebulosity are reliable for most clusters, but the perceived behaviours of individual members can change, so INDICATE should be used as a measure of spatial behaviours between members as a function of their intrinsic properties (e.g., mass, age, object type), rather than to draw conclusions about any specific observed member.}{INDICATE is a robust spatial analysis tool to reliably study Gaia-observed young cluster populations within 1\,kpc, up to a sample incompleteness of $83.3\%$ and binarity of $50\%$.}


\authorrunning{Buckner et al.}

\maketitle 

\section{Introduction}\label{sect_intro}

Young massive clusters (YMCs) are an integral part of the active star formation process in our galaxy, and so have the ability to provide important clues about the formation of massive stars through analyses of  substructure and star--gas dynamics, for example. As such, these clusters have been the focus of intense observational study for decades.  In recent years spatial distribution analyses of stellar members have become a focal point for the community as they give insights into cluster formation histories and early evolution. 

Two complementary types of spatial distribution analysis can be distinguished. The first aims to identify  discrete sub-structures (e.g. sub-clusters, filaments) and to characterise them into topological entity sets (e.g. \citealt{kuhn_spatial_2014}, \citealt{2016MNRAS.456.2900A}, \citealt{2017MNRAS.465.1889G}, \citealt{2019ASPC..523...87J}, \citealt{2021A&A...647A..14G}); the second focuses on characterising the relative positions and spatial behaviours of stars (e.g. \citealt{2015MNRAS.449.3381P}, \citealt{2017A&A...599A..14J}, \citealt{2019MNRAS.490.2521H}, \citealt{2020A&A...636A..80B}). With this second approach the degree of mass segregation, for example, can be obtained from the position of the most massive stars, while insights into the dynamical evolution and star formation imprints can be obtained through a comparison of the positions of the most and least evolved members. 

Prior to the second release of the Gaia survey (DR2; \citealt{2018A&A...616A...1G}) the majority of clusters lacked reliable parallax and/or distance measurements for their suspected members, which were typically identified from photometric analysis alone. With DR2, parallax measurements of unprecedented accuracy became available and an inevitable culling of membership lists ensued \citep{2018A&A...618A..93C}. Unfortunately, this refinement has come at a price. As an optical-wavelength survey, Gaia is highly susceptible to both line-of-sight (LoS) and natal cloud extinction, so significant sample incompleteness in membership lists is unavoidable, even with the additional data provided by the (early) DR3 release. 

The full impact of this incompleteness on the conclusions drawn about clustering properties from spatial distribution studies is unclear, but previous studies have shown that the affects of observational completeness are not trivial (\citealt{2009Ap&SS.324..113A}, \citealt{2012A&A...545A.122P}). An important consideration is the significance (and persistence) of apparent spatial distribution patterns and morphological features for datasets that suffer from such incompleteness. For example, a question arises regarding the identified differences in the spatial behaviour of high- and low-mass stars in a given cluster. It is not clear whether the differences are real or due to a disproportionate number of the lower-mass members being absent from the sample. This is a particular issue with the addition of dynamical data from Gaia,   typically only available for a fraction of the already incomplete sample,  as it could lead to the over-interpretation of identified spatial trends.

Our aim in this paper series is the development of a spatial distribution tool to characterise the relative positions and spatial behaviours of stars, optimised for young stellar cluster analysis. In Paper I we introduce the INdex to Define Inherent Clustering And TEndencies  (INDICATE; \citealt{2019A&A...622A.184B}), which assesses and quantifies the degree of spatial clustering of each object in a dataset, and demonstrated its effectiveness as a tracer of morphological features. In Paper II \citep{2020A&A...636A..80B} we show  that when combined with kinematic data from Gaia DR2, INDICATE is a powerful tool for analysing the star formation history of a cluster in a robust manner. In this paper the impact of incomplete  Gaia-observed datasets on results obtained by INDICATE for clusters is considered. We generated a series of clusters and accompanying synthetic observations of how the clusters would appear through a Gaia filter at various distances and reddening scenarios. INDICATE was applied to each cluster and its corresponding observation, then the results for each were directly compared. 

This paper is structured as follows. In Section\,\ref{sect_clusters} we detail how our synthetic clusters and observations are generated. Our analysis methods are described in Section\,\ref{sect_method} and the results are presented in Section\,\ref{sect_results}. A discussion of these results and our conclusions are given in Section\,\ref{sect_discuss}. Reference tables of expected index changes owing to sample incompleteness as a function of cluster distance, average extinction, binarity, and stellar masses are provided in Appendix\,\ref{app_ref}.

\section{Cluster sample}\label{sect_clusters}

\subsection{Synthetic dataset}

To emulate spatial distributions of many young regions, and ensure that the results of our analysis are statistically representative, we generate ten sets consisting of four synthetic clusters of age 5\,Myr with 300 members   using the McLuster code \citep{2011MNRAS.417.2300K}. Each set draws 300 stars from the canonical \citet{2001MNRAS.322..231K} initial mass function (IMF) with a lower and upper limit of 0.08\,$M_{\odot}$ and 100\,$M_{\odot}$, respectively. These stars are then placed into three spatial configuration realisations of fractal dimension $D=2.0$ to create three of the set clusters. The fourth cluster of the set is a control cluster where we place the stars in a random spatial configuration. We vary the fraction of binaries for the clusters in order to gauge whether this has a significant impact on the results of INDICATE. Each cluster in the set is assigned a binary fraction, $f$, of either 0.0, 0.25, or 0.5 (Sect.\,\ref{sect_binaries}). Table\,\ref{tab_cl_ref} summarises the spatial distributions and binarity used to generate our synthetic cluster dataset. 

\begin{table}
\caption{Summary of the spatial distributions and binarity used to generate our synthetic cluster dataset. }              
\label{tab_cl_ref}      
\centering                                      
\begin{tabular}{c | c |c }          
\hline\hline                        
Distribution & Binarity & Number of Realisations \\    
\hline                                   
    Fractal ($D=2.0$) & $f=0.00$ & 10 \\      
    Fractal ($D=2.0$) & $f=0.25$ & 10 \\
    Fractal ($D=2.0$) & $f=0.50$ & 10 \\
    Random & $f=0.00$ & 10 \\

\hline                                             
\end{tabular}
\end{table}
We refrain from using cluster simulations with a physical underpinning (e.g. evolved using NBody6) as our aim in the current study is to benchmark INDICATE's performance on datasets that are incomplete due to Gaia limitations. For this a statistics-based analysis is essential to confirm that our results are representative and typical of what one can expect when using INDICATE on an incomplete Gaia-observed cluster rather than unique to any single cluster. As such, we only require datasets to approximate the observed spatial distributions of young regions, which is achieved through the above prescription and is significantly less computationally expensive than running, for example, 40 NBody6 simulations through to 5\,Myrs.

We do not explore the effect of number of members, specific spatial distributions ($D$ values), size, or stellar density for clusters as INDICATE is a local statistic that   works independently of these factors \citep{2019A&A...622A.184B}.  

\subsection{Binary set-up}\label{sect_binaries}

For clusters assigned a binary fraction of $f>0$, binaries are created as follows. Primary and secondary binary components are selected automatically by the McLuster code from the 300 member stars already drawn. This selection is made independently of the masses of the two components, resulting in a potential mass ratio between $8\times10^{-4}$ and 1 for pairings (though in practice it is $5\times10^{-3}$ to 1). Separation distances between components range between 0.05 AU and 15105.94 AU, drawn from the \citet{1995MNRAS.277.1491K} period distribution.  For context, the typical distance of single (non-binary) stars to their first nearest neighbour (1-NND) is $\sim$18,000 AU. After each pairing is made, the two component stars are temporarily replaced by a centre-of-mass particle and only reinstated after the cluster's density profile is established and the member velocities scaled. The orientation of the binaries orbital planes, and their orbital phases, are randomly assigned by the code. Binary eccentricity values, $e$, are drawn from a thermal eccentricity distribution ($f(e) = 2e$) and the analytical correction of \citet{1995MNRAS.277.1507K} for the lack of high-eccentricity short-period binaries in the Milky Way applied. Further details on how binaries are set up by McLuster can be found in Appendix A8 of \citet{2011MNRAS.417.2300K}.

Although  binary stars with very wide separations of $>$100,000 AU have been found in the field \citep{2020ApJS..247...66H} and should also exist in young associations, we chose not to include them in our simulations because  INDICATE is a local indicator tool. It describes the spatial distribution in the immediate neighbourhood of a star with an index that is dependent on both the number of neighbours and the separation distance between neighbours, and does not take the wider cluster region into account (see Sect.\,\ref{sect_method_indicate}). For two stars with the same number of neighbours, the star with the smallest separation to its neighbours has the higher index value. Binary stars in pairings at smaller separations than the average 1-NND of single stars in the cluster will hence typically have a higher index value than single stars (assuming a similar number of single neighbours), but the index of binaries in pairings larger than 1-NND should not significantly differ from those of the single star index range in the cluster. Therefore, incompleteness has the potential for a  greater  impact on the derived index values of smaller separation binaries than those of single stars or wider separation binaries; and the contrast separation and resolution limits of Gaia will most strongly impact the detection of smaller separation binaries.  Thus, it is important to ascertain specifically how the perceived spatial behaviour, as seen by INDICATE, of these types of binaries is  affected,  and also the behaviour of  host clusters with large fractions of these binaries. We note that, as far as any INDICATE analysis is concerned, the definition of close and wide binaries is only that the separation is respectively less and more than the typical 1-NND of the region rather than a specific AU value as INDICATE is independent of angular size \citep{2019A&A...622A.184B}. This means that for the index to be potentially affected differently to single stars, the separation distance needs to be less than the 1-NDD of the cluster being studied.

\begin{figure*}
\centering
   \includegraphics[width=0.49\textwidth]{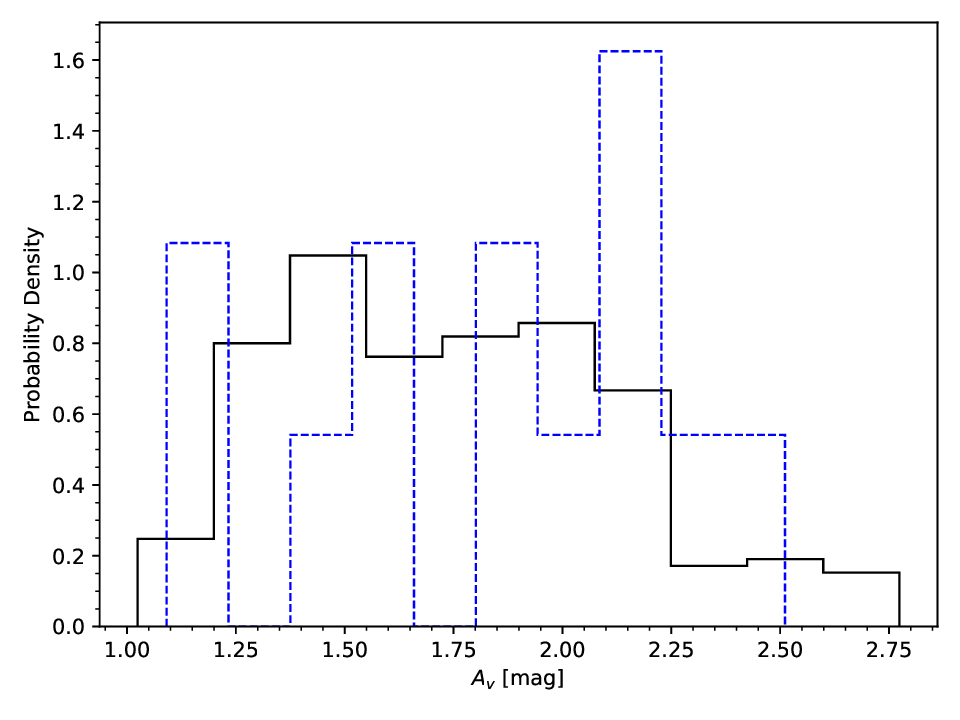}\hfill
   \includegraphics[width=0.49\textwidth]{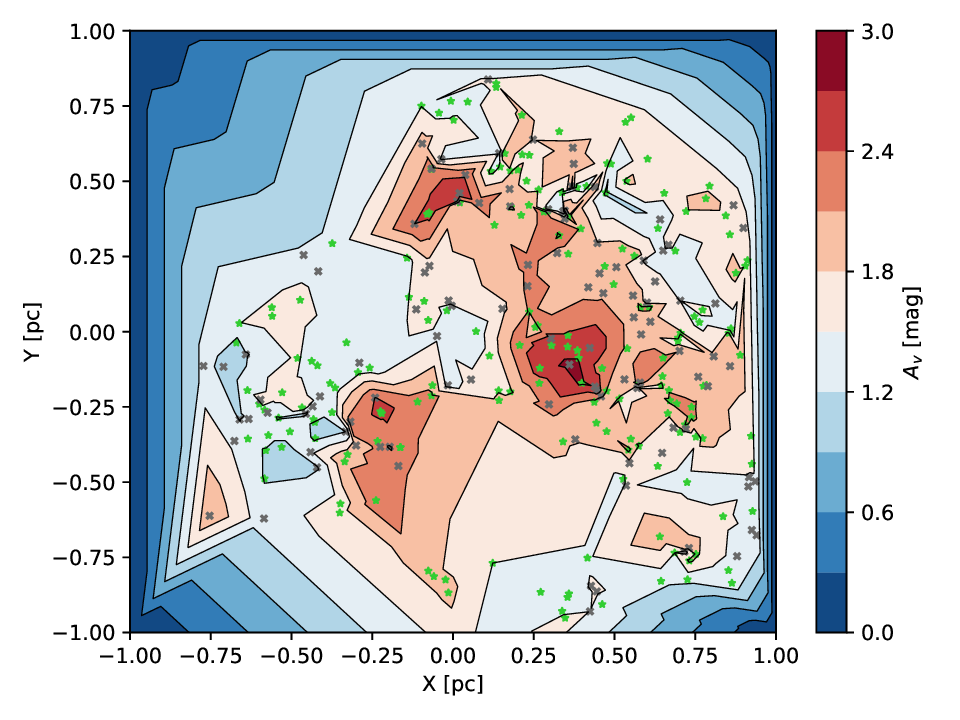}
  \caption{Distribution of variable extinction values generated for members of a cluster with no binaries as part of the Obs-B observations.  Left panel: Normalised histogram of $A_{\text{v}}$ values for OB members (blue dashed line) and all members (solid black line). Right panel: Corresponding visual extinction map with star positions overlaid. Stars observed and not found by Gaia are shown  as green stars and grey crosses, respectively.} \label{fig_AVmap}
\end{figure*}

\subsection{Mock Gaia catalogue}

For each cluster we consider the potential causes of incompleteness when observed by Gaia. One cause is the apparent magnitude of members outside Gaia’s detection limits owing to cluster distance, uniform extinction, and/or variable extinction. Another factor is the contrast and sensitivity limitations as a function of angular separation and flux ratio (magnitude difference).  

To  produce synthetic Gaia observations of the clusters we use the Gaia simulator{\footnote{\url{https://github.com/zkhorrami/gaiaSimulations}}} tool which is part of the larger Make Your Own Synthetic ObservaTIonS{\footnote{\url{https://github.com/zkhorrami/MYOSOTIS}}} (MYOSOTIS; \citealt{2019MNRAS.485.3124K}) tool. Given the stars' position, age, mass, and extinction values (or gas cloud), MYOSOTIS estimates their synthetic Gaia magnitudes in the desired filter (G, GBp, GRp), using stellar evolutionary and atmosphere models. We used the Dmodel extinction model of MYOSOTIS with R$_V=3.1$ and the solar metallicity (Z=0.015) for simulating stellar fluxes. The OBtreatment option is also set for high-mass stars ($T_{eff} > 15$ kK) so that proper spectral energy distributions (SEDs) were chosen to estimate the flux of hot O- and B-type stars.

To determine the impact of incompleteness owing to cluster distance and LoS extinction across the cluster region, we instructed MYOSOTIS to produce two sets of simulations for the cluster catalogue, adopting different techniques to apply extinction for stars within the clusters. First, Obs-A for which  a constant $A_V$ value is applied to all stars within the clusters (Sect.\,\ref{sec_obA}), then Obs-B for which variable extinction is appiled across the field of view by positioning each cluster at the centre of a homogeneous smoothed particle hydrodynamics (SPH)  gas cloud (Sect.\,\ref{sec_obB}, \ref{sec_exmaps}).

\subsubsection{Obs-A}\label{sec_obA}

Obs-A are observations for each cluster at a distance of $100\,\text{pc}\,\le\,d\,\le\,1000\,\text{pc}$ in 100\,pc increments with a constant extinction value of  $0\,\text{mag}\,\le\,A_{\text{v}}\,\le\,16\,\text{mag}$  in 1\,mag increments for every member star. This resulted in 10 x 17 = 170 observations for each cluster, and 6800 total observations. The maximum LoS extinction was set at 16\,mag for two reasons. First,  this is six magnitudes higher than the highest value for a cluster within 1\,kpc in the MWSC Catalogue \citep{2013A&A...558A..53K} that is not associated with natal nebulosity, so it should be sufficient to cover values of galactic clusters discovered in the future;  second, when the clusters were given extinction values higher than 16\,mag, the number of members detected by Gaia was consistently less than 50 (INDICATE's minimum sample size limit is 50; \citealt{2019A&A...622A.184B}). 

\subsubsection{Obs-B}\label{sec_obB}

Obs-B are observations with variable extinction across the cluster region. The aim of these tests is to appraise how well INDICATE handles the ‘patchy’ incompleteness associated with young embedded clusters. We do not attempt to produce a realistic physical approximation of a natal cluster environment or a specific observed region, but rather plausible extinction maps for the clusters (Sect.\,\ref{sec_exmaps}). As the affect of cluster distance and uniform foreground extinction on the index was explored in Obs-A, we keep these constant in Obs-B to ensure that any found changes are attributable to the spatially heterogenous incompleteness caused by the variable extinction typically associated with these regions. One observation of each synthetic cluster was made, resulting in 40 total observations.

\subsubsection{Extinction map set-up}\label{sec_exmaps}

To create the maps, each cluster is observed at 1\,kpc with no foreground extinction inside a uniformly composed spherical   SPH cloud with no turbulence or structure. After careful consideration of the parameters, our cloud consists of $10^5$ particles (total mass $2\times 10^3 M_{\odot}$), a radius  at least three times larger than the cluster's radius, and a gas column density of $N_H=3.28\times 10^{21} [cm^{-2}]$. The resulting cloud produces A$_V$ values within the range of 0 mag (for a foreground star) up to 3.18 mag (for a background star), and 1.48\,mag at the cloud's centre. Each cluster is placed centrally inside the cloud, such that each star’s extinction value is determined by its Z-axis position: the deeper the Z-axis position, the greater the LoS cloud depth, the greater its A$_V$. MYOSOTIS solves the RT equation for each star, accounting for the extinction provided by the SPH particles  (see \citealt{2019MNRAS.485.3124K} for full details). The result is varied extinction across the cluster, as shown in the right panel of Figure\,\ref{fig_AVmap}.

We note that the aim of the Obs-B tests is to appraise how well INDICATE handles spatially heterogeneous incompleteness, so a realistic physical approximation of a natal cluster environment is not required for these purposes. INDICATE  has already been shown to produce robust results for embedded clusters where such structure and incompleteness is present \citep{2020A&A...636A..80B}. Our aim in this work is to generalise this result to provide quantitative guidance for users of INDICATE regarding the reliability of the index values as a true reflection of spatial behaviours (rather than observational biases) in these regions, and thus no structure (e.g. gas clumps, filaments) was included in the cloud simulation.  

The presence of structure and its potential to contribute to sample incompleteness is not trivial. For example, depending on the physical scale of the structure, it is reasonable to expect large regions of a cluster and/or companions from stellar pairs in some cases (but perhaps not all)  to be obscured. It is also possible for such structure to exist in a region and have a minimal impact on incompleteness due to the relative position of the stellar population. Physically each region has unique stellar and structural spatial patterns, so the exact pattern and extent of incompleteness will vary from region to region. However, statistically, this remains a patchy incompleteness problem (i.e. stars are removed heterogeneously from the dataset). Adding structure to our cloud only specifies the exact locations of that incompleteness. Thus, as our aim is to assess INDICATE's general ability to handle this type of incompleteness, the mechanism of the incompleteness is less important than the result; in other words (i) each cluster has a realistic range of stellar extinction values and subsequent likelihoods of detection by Gaia and (ii) variation in pattern and extent of the incompleteness between each observed cluster. To ensure stellar extinction values that can reasonably be expected to be observed, we carefully chose the parameters of the cloud so that the resulting stellar extinctions have a similar dispersion to those typically found in galactic young open star clusters (\citet{2017PASA...34...68R} and Figure\,\ref{fig_AVmap} left panel). As the 3D stellar positions of each of the 40 clusters in the test is unique (i.e. no two clusters are spatially identical), the exact pattern and extent of incompleteness varies from cluster to cluster.  

The generalised results presented in this study provide sufficient knowledge of the index’s behaviour to aid interpretation of significant values of observed clusters in regions of variable extinction, but if its specific behaviour in any given region is desired we recommend that users run the Obs-B tests again with that region’s observed extinction map.\\

\subsection{Resolution limitations}

For each synthetic observation, stars with an apparent magnitude outside the sensitivity limit of Gaia ($3\,\text{mag}\,\le\,G\,\le\,21\,\text{mag}$) are removed. We assume all stars in the observations to be true members and that there is no field star contamination (as this issue, and its impact on INDICATE, is addressed in \citealt{2019A&A...622A.184B}). However we remain mindful that in most cases, even with the best of efforts, not all field stars will be removed from observationally obtained datasets prior to analysis. This is reflected in our choice of $N=5$ for INDICATE's nearest neighbour number in this work(see Sect. \ref{sect_method_indicate}). As discussed in \citet{2019A&A...622A.184B}, while the index values of true cluster members are generally unaffected by the presence of interloping field stars, the proportion with an error (deviation from their true value), and the size of that error, scales with increasing nearest neighbour number and level of contamination, reaching a maximum of  $\sim\,95\%$ of members having a non-zero error with $100\%$ uniform field star contamination and $N= 9$ (a similar effect is found when field stars are distributed as a gradient). Therefore, it is desirable to use a small value of $N$ when field stars may be present, but as $N$ essentially defines the resolution, a value that is  not too small should
be chosen so that  subtle (larger-scale) clustering tendencies are not missed. As demonstrated in \citet{2020A&A...636A..80B}, a value of $N=5$ strikes a good balance between these two considerations and  produces robust results for observed clusters. 

We also apply known contrast sensitivity limitations as a function of angular separation and magnitude difference at the 99\% detection level \citep{2019A&A...621A..86B}, and remove stars that are unresolved by Gaia. This detection threshold was selected because it is the harshest, and  therefore will result in more incomplete datasets,  and because datasets will consist of stars only with a high likelihood of detection, thus enabling us to explore INDICATE's ability to analyse the worse case scenarios in Gaia cluster catalogues. Contrast limitation is most important for visually close stellar systems and for clusters with binaries, as a bright star can mask a companion and/or close neighbours depending on their respective fluxes and separation. In real Gaia observations this masking effect will, in some cases, cause the companion to vanish (i.e. be removed from the catalogue), but in other cases Gaia will detect the combined light from both components, making the primary appear as a single overluminous source. In all cases we justify removing a masked companion or neighbour  from our synthetic observations  (rather than combining its flux with the primary) as INDICATE only requires the number of stars detected and their spatial positions to calculate stellar index values;  stellar flux values are not utilised. Therefore, spatially, cases for which pairs are combined into a single source are equivalent to the removal of masked companions from the catalogue. In the scenario a combined flux would have resulted in an intermediate-mass star appearing to be a high-mass star, this should have a negligible affect on the observed spatial properties of the high-mass population. If this population exhibits overall different spatial tendencies to the low- and intermediate-mass stars, (i) INDICATE is robust against outliers \citep{2019A&A...622A.184B} and (ii) stars masquerading as another class (e.g. intermediate-mass as high-mass) are easily identified by their index value, which  will notably differ from the index values of real members of that class  \citep{2020A&A...636A..80B}. 

\begin{figure*}
\centering
   \includegraphics[width=0.33\textwidth]{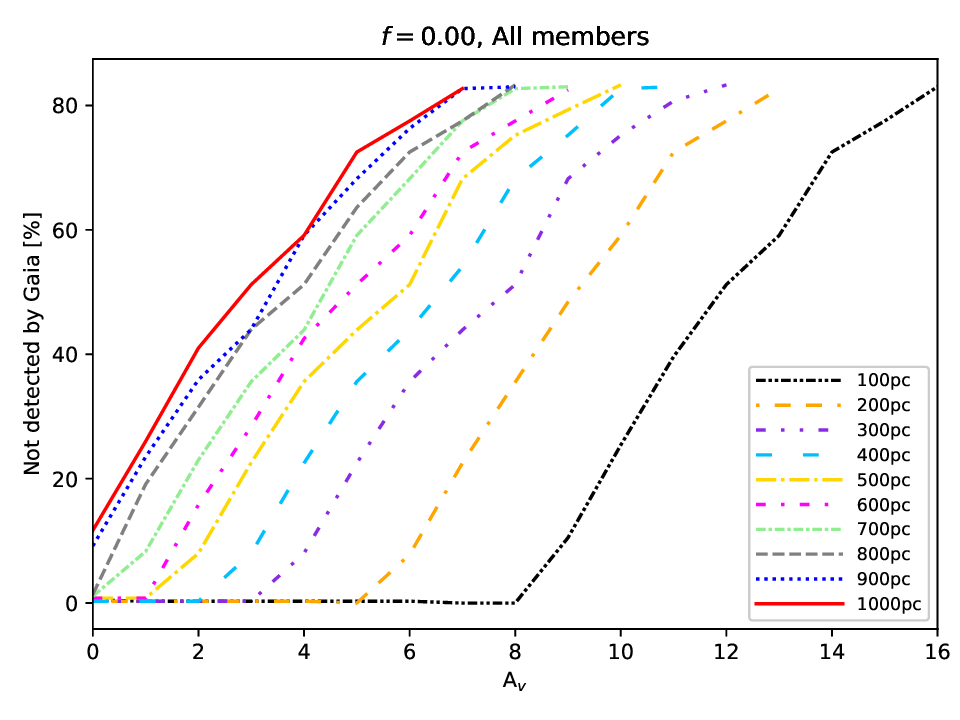}
      \includegraphics[width=0.33\textwidth]{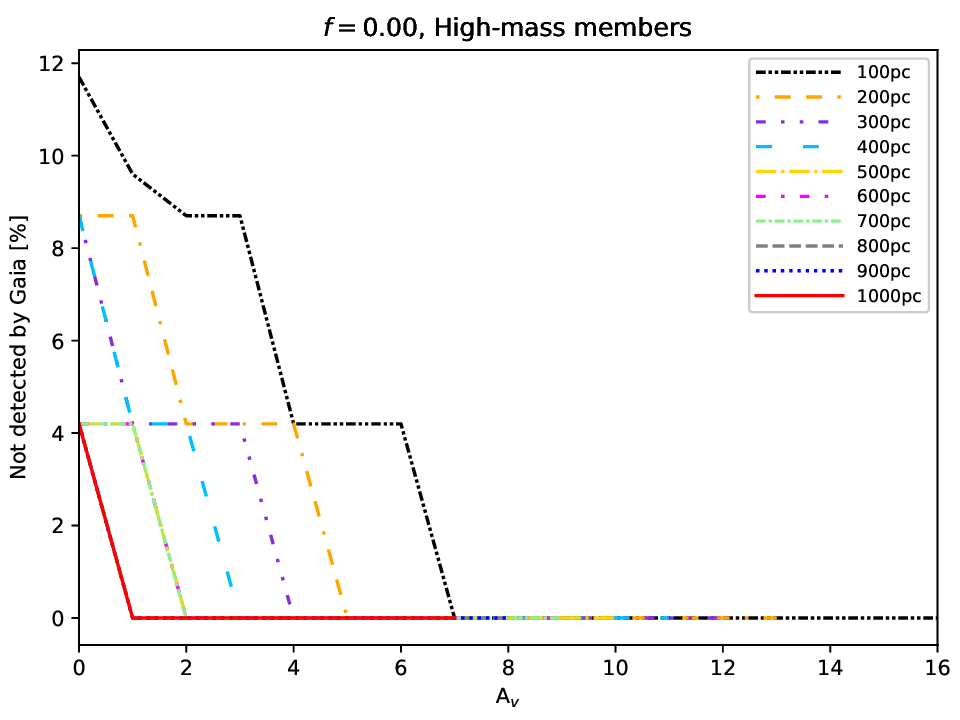}
   \includegraphics[width=0.33\textwidth]{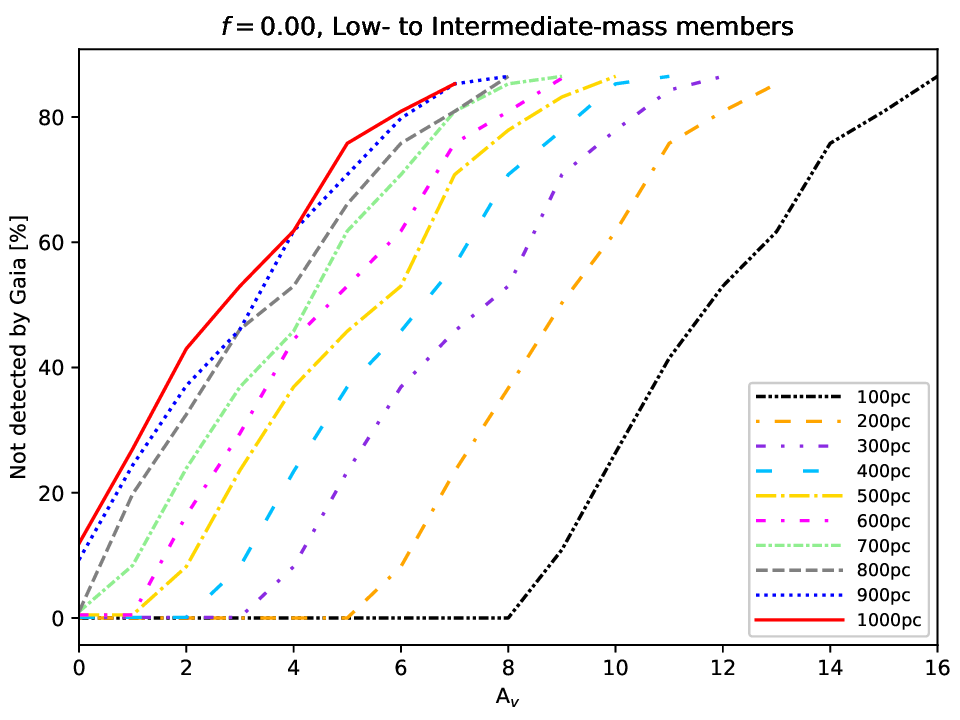}
   \includegraphics[width=0.33\textwidth]{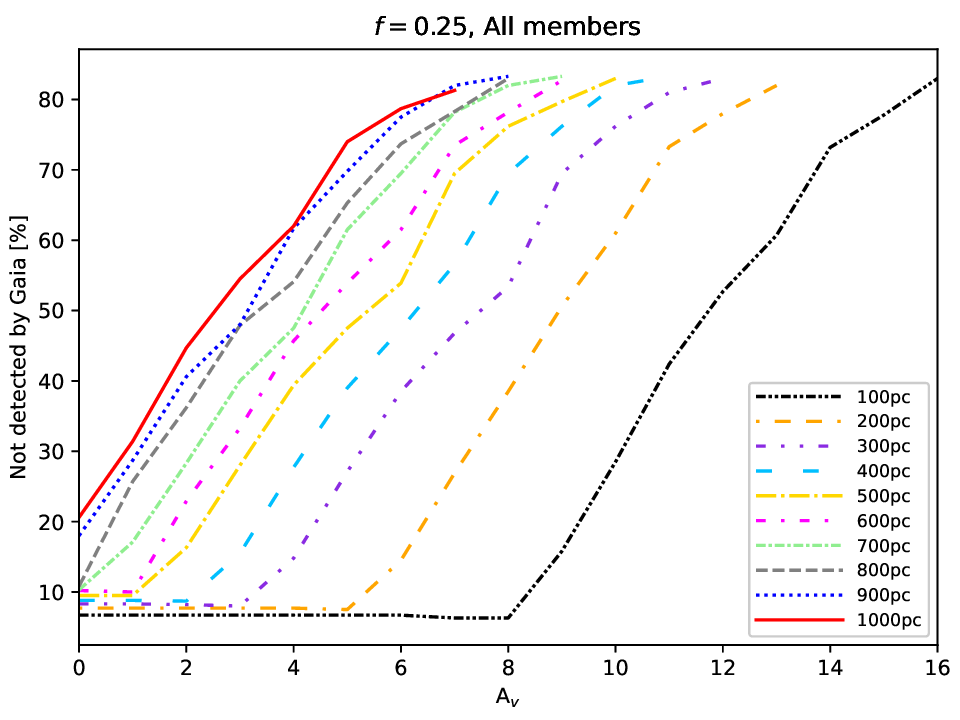}
      \includegraphics[width=0.33\textwidth]{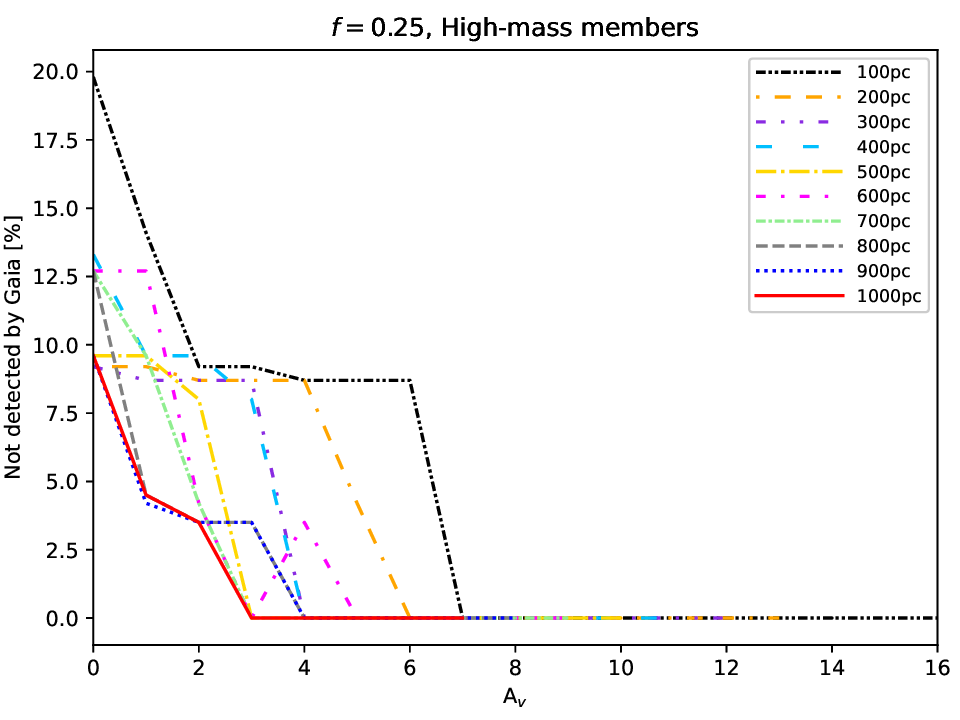}
   \includegraphics[width=0.33\textwidth]{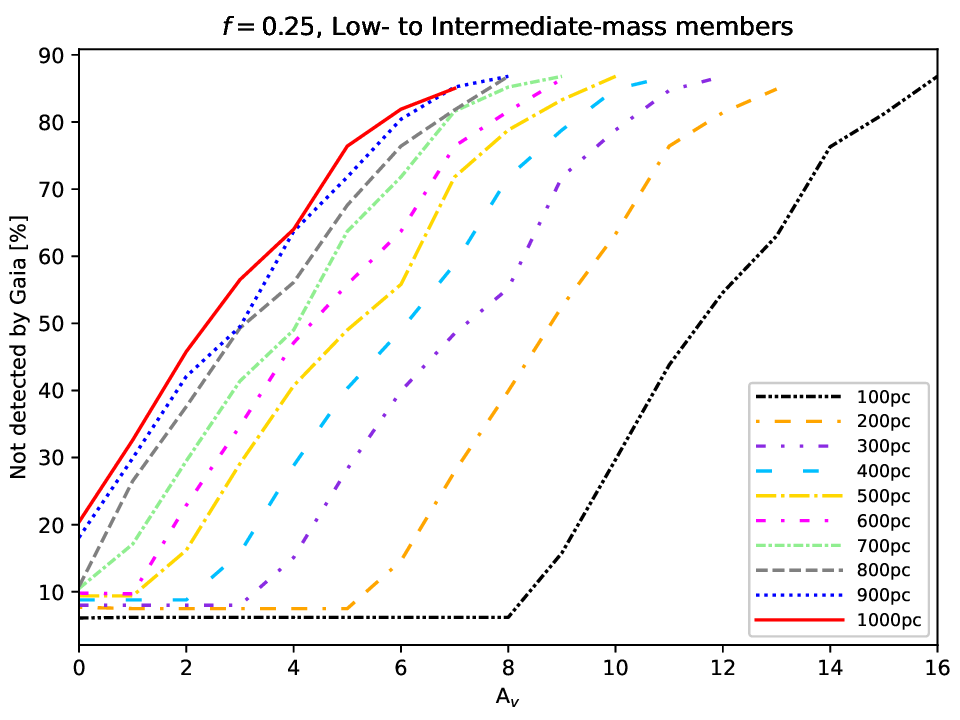}
      \includegraphics[width=0.33\textwidth]{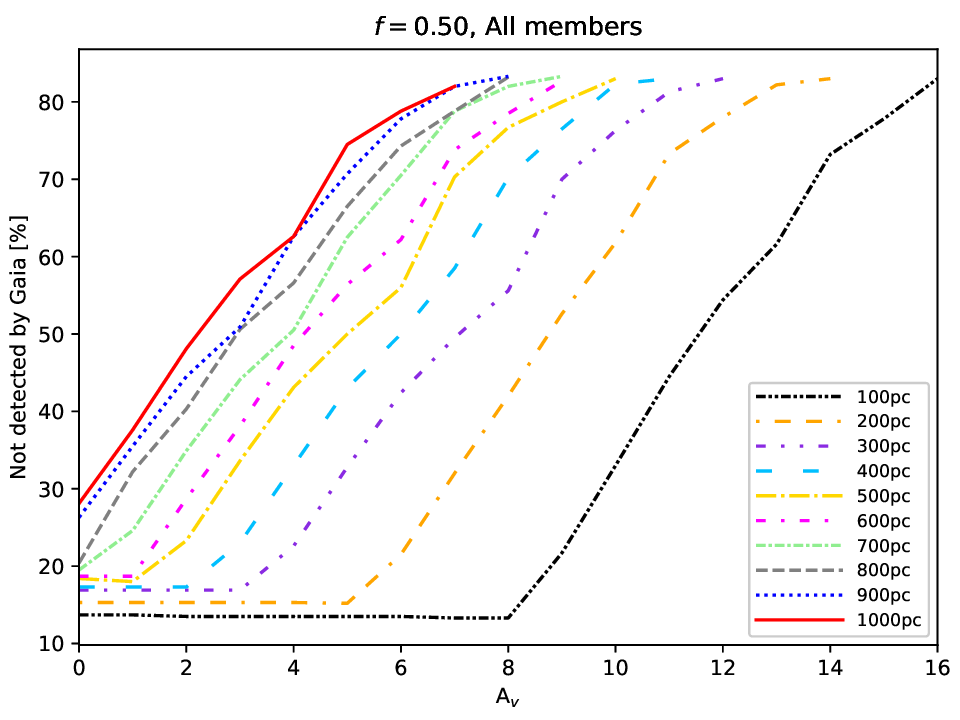}
      \includegraphics[width=0.33\textwidth]{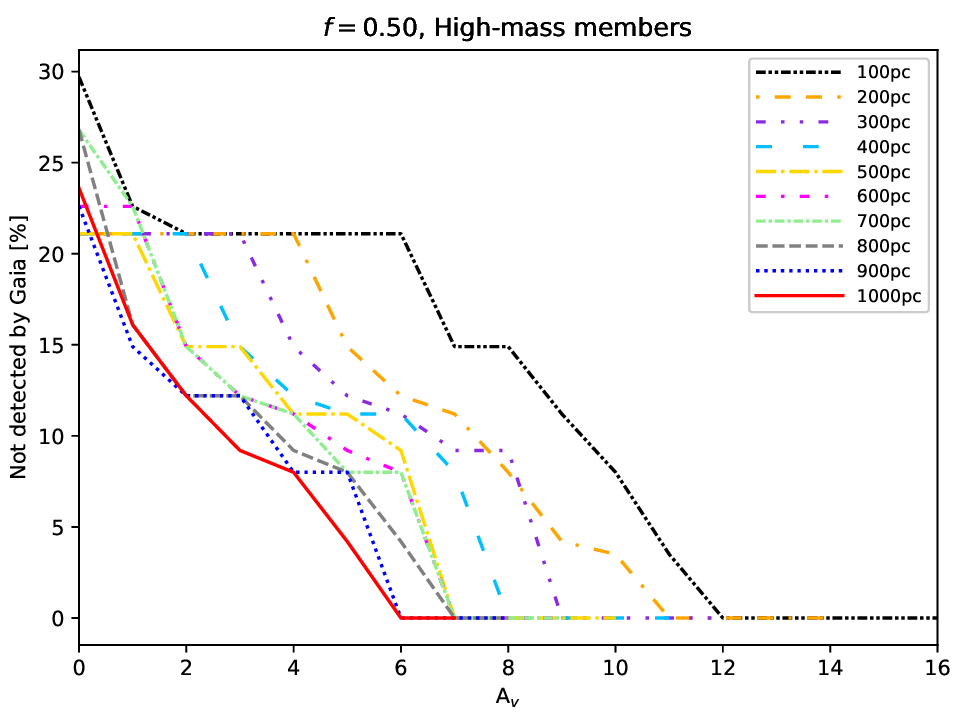}
   \includegraphics[width=0.33\textwidth]{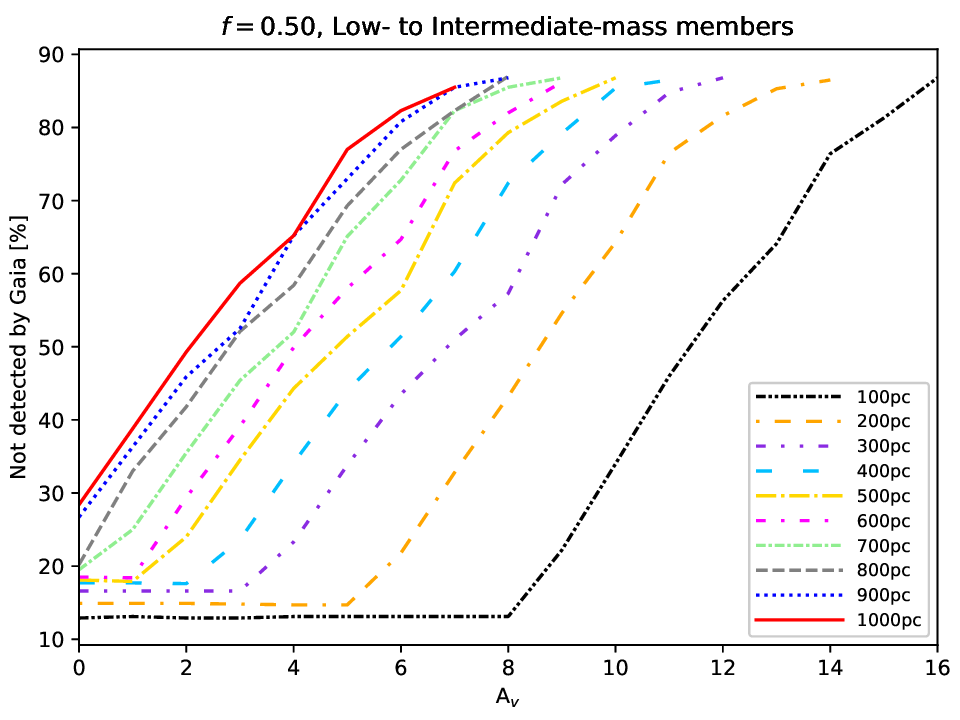}
   \caption{Shows for the (Left column:) general, (Middle column:) high-mass, (Right column:) low- to intermediate-mass population completeness as a function of cluster distance, $A_{v}$ and binarity, for clusters with a binary fraction of (Top row:) $f=0.0$, (Middle row:) $f=0.25$, (Bottom row:) $f=0.50$.}\label{fig_gaia}
\end{figure*}


\section{Analysis method}\label{sect_method}

\subsection{INDICATE}\label{sect_method_indicate}

\citet{2019A&A...622A.184B} introduced the statistical spatial analysis tool INDICATE,\footnote{\url{https://github.com/abuckner89/INDICATE}} which quantifies the degree of association in a cluster by deriving and assigning an index value for each star individually rather than a single value to the cluster as a whole.  

The index is defined as follows. For a cluster of size $n$, an evenly spaced uniform (i.e. definitively non-clustered) control distribution of the same density is generated across the parameter space. The mean Euclidean distance, $\bar{r}$, of every star $j$ in the cluster to its fifth-nearest neighbour in the control is measured, and its index value is calculated as

        \begin{equation} \label{eq_I}
                \\ I_{5,j}= \frac{N_{\bar{r}}}{5}
        ,\end{equation}where $N_{\bar{r}}$ is the number of actual nearest neighbours to star $j$ within a radius of $\bar{r}$ in the cluster. The index value $I_{5,j}$ is unit-less with a value range of $0 \le I_{5,j} \le \frac{n-1}{5}$ and the higher the value, the more tightly clustered a star is. 

To determine if a star is spatially clustered (rather than randomly distributed), the index is calibrated. For each application 100 realisations of a random distribution of cluster size $n$ are generated in the same parameter space as the dataset, INDICATE applied, and the mean index values of all random distributions, $\bar{I_5}^{random}$, determined. Star $j$ is then considered clustered if it has an index value above a significance threshold, $I_{sig}$, of three standard deviations, $\sigma$, above $\bar{I_5}^{random}$, i.e. 

        \begin{equation}\label{eq_IjIsig}
        \\ I_{5,\,j}> I_{sig} \text{\,,\,\,\,\,\,where\,\,\,\,\,} I_{sig}=\bar{I_5}^{random}+3\sigma
        .\end{equation}

Using this definition, $99.7\%$ of stars that are distributed in a spatially random configuration will have an index value of $I_{5}<I_{sig}$. Extensive statistical testing by the authors has shown the index to be robust against outliers and edge effects, and there is no dependence between the index and a cluster's shape, size, or stellar density (see \citealt{2019A&A...622A.184B} for a discussion). 

As INDICATE is valid for sample sizes of 50 and greater, we remove any cluster from our catalogue that is observed to have fewer than 50 stars, subsequently reducing the total number of Obs-A observations from 6800 to 4184. We note that this minimum sample size limit caps the maximum cluster member incompleteness permitted in this study at ($250/300=$) $83.3\%$.

\subsection{Statistical considerations}\label{sect_method_stats}

To ensure that the reported changes to the index values in a cluster are representative of,  
 and typical for, the stated observational conditions, we report the average changes from pre- to post-observation in our analysis for clusters with the same binary fraction and observing conditions. This is to compensate for small statistical variations owing to differences in the realisations of the spatial distribution of stars. For example, in each set there is a cluster with a binary fraction of $50\%$, which has been observed at 100\,pc, and has a uniform extinction of $A_{\text{v}}=1$\,mag. Therefore, as there are ten cluster sets, there are ten independent observations of a cluster with [$f=0.5$, 100\,pc, $A_{\text{v}}=1$\,mag], so the values quoted for these conditions are an average derived from the ten realisations.

\begin{table}
\caption{Summary of average INDICATE values for our synthetic fractal cluster datasets.}              
\label{table_Isig}      
\centering                                      
\begin{tabular}{c|c|c|c|c}          
\hline\hline                        
Binarity& Population& $\tilde{I}_{5}$ & $\tilde{I}^{cl}_{5}$ & $max(I_{5})$ \\ 
\hline 
$f=0.00$ &  All &  1.4&3.0 & 4.8 \\
$f=0.25$ &  All &  1.6& 3.0& 5.8 \\
$f=0.50$ &  All &  2.0& 3.2& 7.8 \\
\hline
$f=0.00$ & High-mass & 1.4 & 2.8& 4.4 \\
$f=0.25$ & High-mass & 1.6 &2.7&  5.4\\
$f=0.50$ &  High-mass &  2.0& 3.2&  5.4\\
\hline
$f=0.00$ & Low- to intermediate-mass  &  1.4& 3.0& 4.8 \\
$f=0.25$ & Low- to intermediate-mass  & 1.6 & 3.0& 5.8  \\
$f=0.50$ & Low- to intermediate-mass  & 1.8 & 3.2& 7.8  \\

\hline 

\end{tabular}
\end{table}

\section{Results}\label{sect_results}

\begin{figure}
\centering
   \includegraphics[width=0.5\textwidth]{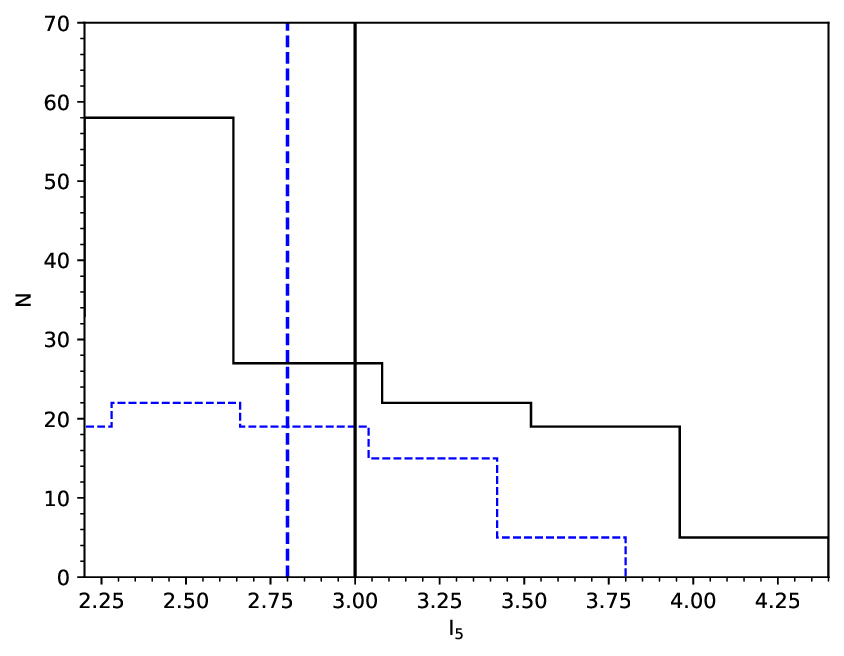}
  \caption{Distribution of member index values above the significance threshold for the cluster shown in Fig\,\ref{fig_AVmap}. Pre-observed distribution and median values are represented by the black solid lines, and post-observed values by the blue dashed lines.}  \label{fig_Ihist}
\end{figure}

In this section we describe the changes in the perceived 2D spatial behaviour of the clusters,  due to the conditions they are observed under, through comparison of INDICATE’s index values pre- and post-observation. Pre-observed cluster values are listed in Table\,\ref{table_Isig}, and an example histogram of the index values for a cluster derived pre- to post-observation is shown in Figure\,\ref{fig_Ihist}.

For Obs-A we provide reference Tables\,\ref{table_b0_all}-\ref{table_b50_Nob} for typical index changes as a function of observed cluster distance and mean $A_{v}$. Figure\,\ref{fig_gaia} shows the sample completeness of the general, high-mass, and low- to intermediate-mass populations as a function of cluster distance, extinction and binarity of these observations.

\subsection{General spatial properties}

\subsubsection{Obs-A} \label{general_results_a}

As expected, the proportion of absent members increases as a function of increasing distance and extinction. The minimum number of absent members is dependant on binarity, starting from $0\%$ ($f=0.0$), $6.3\%$  ($f=0.25$), and $13.3\%$ ($f=0.5$) for low-distance clusters. This reflects the contrast separation distance sensitivity limitations of Gaia as binary members typically have smaller angular separations than unpaired neighbouring members, so dimmer companions are not detected. Similar to the fractal clusters with no binaries, all members of nearby low-extinction clusters in a random configuration (which also have no binaries) are detected, thereby confirming that Gaia member detection is not dependent on the clusters’ fractal dimension. 

For resolved members there is a correlation between increasing binary and completeness with decreasing index value pre- and post-observation. The proportion of members identified as spatially clustered typically decreases by less than 10 percentage points and no more than 26 percentage points for highly incomplete membership lists. As shown in Figure\,\ref{fig_Ipercent} the perceived spatial behaviour of members identified as spatially clustered pre- and post-observation typically decreases by $< 20\%$ even when $83.3\%$ of members are not resolved, and no dependence on binary or spatial configuration is found for this group.

\subsubsection{Obs-B}\label{general_results_b}

The number of missing members increases incrementally with binarity, from $38\%$ ($f=0.0$) to $47\%$ ($f=0.5$), and those absent are almost exclusively low- to intermediate-mass stars. Again, clusters in a random spatial configuration have the same degree of incompleteness as fractal clusters with no binaries. 

The proportion of detected members found to be spatially clustered post-observation typically decreases by less than 10 percentage points from pre-observed levels. There are no large changes pre- and post-observation in either the general population’s index values or in those of members identified as spatially clustered, with their median value decreasing by $< 12.5\%$ and $< 6.9\%$, respectively, irrespective of binarity. The perceived spatial behaviour of cluster populations therefore remains largely unchanged despite significant incompleteness. However, individual indices of stars in spatial concentrations can change by up to $100\%$ from their pre-observed to post-observed value. Therefore, when there is variable extinction across a cluster, the index should be used as a measure for trends in spatial behaviour within the population as a function of  object class, age, mass, for example, rather than comparisons of any two individual stars whose observed index values may have been affected to different degrees from their pre-observed values.

No change in the perceived spatial behaviour of members pre- and post-observation in clusters with a random spatial configuration is found.

\subsection{Spatial properties of OB populations}

Mass segregation is a term often used in the literature to describe two quite different spatial realisations. The classic definition refers to the concentration of high-mass stars together at the centre of the host cluster, so can be found by  examination of the radial distribution of members as a function of stellar mass or by calculating the average nearest neighbour distance between high-mass members and comparing it to those between low- to intermediate-mass members (the former is shorter when mass segregation is present; \citealt{2008AJ....135..173S}, \citealt{allison_using_2009},  \citealt{2015MNRAS.449.3381P}).  A somewhat newer definition refers to the concentration of low- and intermediate-mass members around high-mass members (and high-mass members are not required to be concentrated together), so can be found by calculating the average number of nearest neighbours for high-mass members and comparing it to that for low- to intermediate-mass members (the former is higher when mass segregation is present; \citealt{2011MNRAS.416..541M}) . As INDICATE assigns an index to each star, and this value represents the strength of the stellar concentration in a star's immediate neighbourhood, the tool by definition provides a measure of the newer definition of mass segregation (\citealt{2019A&A...622A.184B}, \citealt{george_indicate_paper}). Below we report on the perceived changes to signatures of this type of mass segregation as found by INDICATE within clusters observed by Gaia.

\subsubsection{Obs-A}

In most observations the full OB population is resolved as their apparent magnitudes do not fall below the Gaia sensitivity limit \footnote{Observations where the apparent magnitudes of OB stars  are fainter than the sensitivity limit were excluded from our analysis as these clusters also had fewer than 50 members detected.}. However, for a few cluster observations the most massive OB members are absent as their apparent magnitudes are brighter than Gaia's sensitivity limit. The number of absent OB members is dependant on extinction and distance (see Tables\,\ref{table_b0_ob}, \ref{table_b25_ob}, \ref{table_b50_ob}), but also binarity, such that a maximum of $11.7\%$ ($f=0.0$), $19.8\%$  ($f=0.25$), and $29.7\%$ ($f=0.5$) of the OB population is absent in some observations. This correlation is due to the increased effect of masking by the PSF wings of high-mass stars on their neighbouring stars (including those that are themselves high-mass). As binary members typically have smaller angular separations than unpaired neighbouring members, the impact of PSFs on completeness scales with cluster binarity. Clusters that have a random configuration and no binaries also have a maximum of $11.7\%$ of the OB population absent, which is consistent with our previous result (Sect.\,\ref{general_results_a}) that member detection is independent of the spatial configuration of clusters, due to the high angular resolution achieved by Gaia.

\begin{figure*}
\centering
   \includegraphics[width=0.55\textwidth]{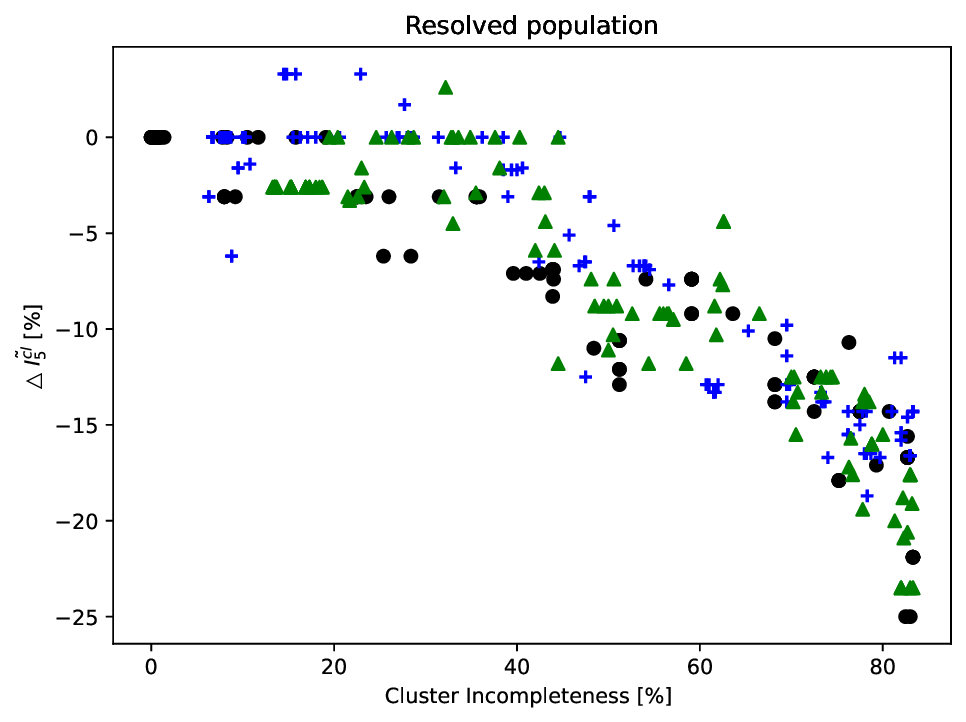}
      \includegraphics[width=0.55\textwidth]{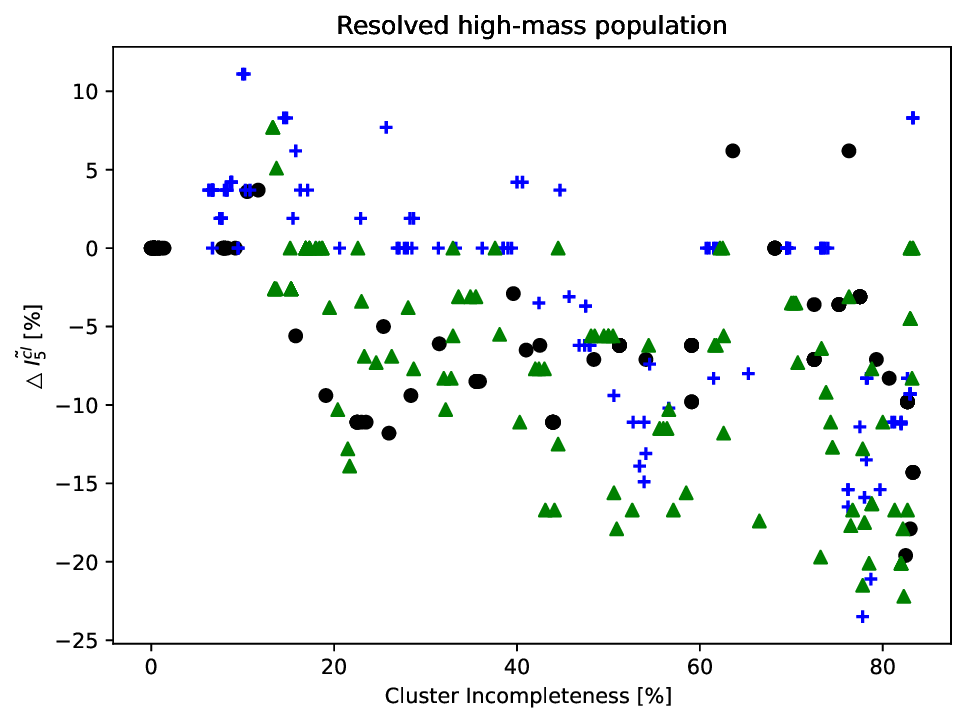}
   \includegraphics[width=0.55\textwidth]{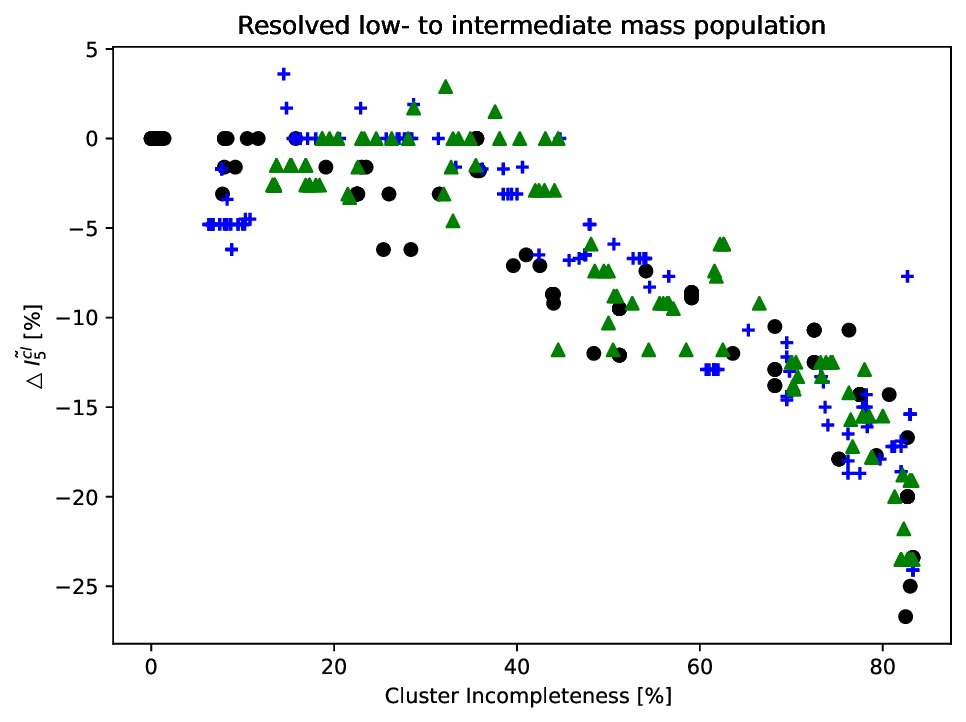}
   \caption{Plot of the change in the median index value derived for stars identified as spatially clustered, $\tilde{I}^{cl}_5$ , in the resolved populations:
 general (top),  high-mass (middle), and  low- to intermediate-mass  (bottom). Symbols and colours indicate the binary fraction of the host cluster: $f = 0.0$ (black circles), $f = 0.25$ (blue crosses), $f = 0.5$ (green triangles). }\label{fig_Ipercent}
\end{figure*}

Typically, there is a decrease between the pre- and post- observed index values of the OB population, and signatures tend to be weaker in clusters with binaries. The proportion of observed OB members found to be clustered ($I_5 > I_{sig}$) post-observation typically decreases by less than 10 percentage points from the pre-observed levels, and by no more than 23 percentage points. A change in the median index values of the clustered members between $+5\%$ and $-20\%$ is found in most clusters, but can decrease up to $36.1\%$ when the degree of completeness is extremely low (Figure\,\ref{fig_Ipercent}). No mass segregation was found in the clusters that have a random distribution, meaning that  INDICATE does not find false signatures of mass segregation in Gaia-observed clusters due to incompleteness bias.

\subsubsection{Obs-B}

Most OB stars are resolved by Gaia, but similarly to Obs-A there is a correlation between the proportion of unresolved members and cluster binarity. All OB stars are resolved in clusters with no binaries, but for clusters with binary fractions of $f = 0.25$ and $f = 0.5$ there is a $4\%$ and $12.5\%$ decrease in resolved OB members. The number of OB stars identified as clustered ($I_{5} > I_{sig}$ ) is independent of binarity and can decrease by 15 percentage points, resulting in a corresponding decrease of up to $14.6\%$ in the median index for these stars compared to pre-observed levels (i.e. index values of high-mass members typically remain unchanged with respect to pre- observed levels), but in some cases they are underestimated.

The full OB population is detected in clusters that have a random distribution. INDICATE correctly determines that no OB stars are clustered in the observations of these clusters.

\subsection{Spatial properties of low- to intermediate-mass populations}

\subsubsection{Obs-A}\label{sect_A_lowmass}

Lower-mass members are the primary source of incompleteness in clusters because they are  intrinsically fainter than their higher-mass counterparts. The proportion of these stars that are unresolved is a function of distance, extinction, and binarity, with minimum absences of $0\%$ ($f=0.0$) to $6.1\%$ ($f=0.25$) and $12.9\%$ ($f=0.5$)  (see Tables\,\ref{table_b0_Nob}, \ref{table_b25_Nob}, \ref{table_b50_Nob} for full details). Again, the correlation between degree of completeness and binary fraction is attributed to the contrast separation distance sensitivity limitations of Gaia.

There is a correlation between increasing binarity and completeness with decreasing index values pre- to post-observation of resolved members. The relative proportion of this population identified as spatially clustered typically decreases pre- to post-observation by less than 10 percentage points , but no more than 27 percentage points  for the most incomplete membership lists. As shown in Figure\,\ref{fig_Ipercent} the perceived spatial behaviour of members identified as spatially clustered pre- and post-observation decreases by $<25\%$ when $83.3\%$ of members are not resolved. These results resemble those of the general population (Sect. \,\ref{general_results_a}) as the lower-mass stars are its primary constituent, and confirms these conclusions regarding spatial behaviour of members pre- and post-observation. A notable change does occur in the highest index value obtained for this population, decreasing pre- to post-observation by up to $\Delta$ max $I_5$ = $62.5\%$ ($f = 0.0$), $60.7\%$ ($f = 0.25$), and $57.6\%$ ($f = 0.5$), which means that  stars in strong spatial concentrations can appear significantly less clustered.

Echoing the general population results, no change in the perceived spatial behaviour of lower-mass members in clusters in a random spatial configurations is found.

\subsubsection{Obs-B}

Lower-mass stars are the primary source of incompleteness in Obs-B. There are no large changes pre- and post-observation in  the population’s index values or in those of members identified as spatially clustered, with their median value decreasing by $< 12.5\%$ and $< 6.5\%$ respectively. The fraction of stars found to be spatially clustered decreases by less than 10 percentage points, and the indices of stars in spatial concentrations change up to $100\%$ from their pre-observed values,  the same as for the general population (Sect.\,\ref{general_results_b}).

No change in the perceived clustering behaviour of lower-mass members in clusters with a random spatial configuration is found.

\section{Discussion and conclusions}\label{sect_discuss}

We summarise the results of our analysis as follows. The 2D spatial behaviours identified by INDICATE are reliable within 1\,kpc for incomplete Gaia-observed datasets of clusters, those   associated with natal nebulosity and those not associated with natal nebulosity. Typically there are no fundamental changes in the conclusions drawn regarding the spatial behaviour of stellar populations from index values obtained pre- and post-observation, but in some clusters the observed strength of stellar associations may be diluted. Most notably, index values did not significantly increase when observed nor were clustering behaviours found to be present in clusters where none actually existed: spatial behaviours identified by INDICATE for cluster populations are real and not generated by observational biases. The perceived behaviours of individual members of the population, however, can be affected so the index should not be used to draw conclusions about any specific member, but rather used as a measure of spatial behaviours between members as a function of their  mass, age, and  object class, for example. In agreement with \citet{2019A&A...622A.184B} we find the spatial configuration of a cluster (fractal, random) to have no influence on INDICATE's index. 

These results were expected as INDICATE is a local statistic, and thus derives the index value of a star by looking only around its immediate neighbourhood rather than at the cluster as a whole. Therefore, (i) the shape of the cluster is not considered when the index is calculated, (ii) the index of stars for which the majority of their neighbours are not resolved will significantly decrease, but (iii) those that are in high spatial concentrations remain in (relatively) strong concentrations even when some neighbours are removed so the effect on their (and the overall population’s) index values is small. For high-mass stars in strong concentrations we find the proportion typically decreases by less than 10 percentage points and a change between $+5\%$ and $-20\%$ in their pre- and post-observed index values occurs, meaning that the conclusions regarding whether a cluster is mass segregated using INDICATE’s index are robust, in contrast with some other methods (e.g. Group segregation ratio, \citealt{2015MNRAS.449.3381P}). We note that this result is valid for typical young star forming regions (as high-mass stars make up a fraction of the observed members and the datasets include some resolved lower-mass neighbours), but may differ for very incomplete clusters with a large population of high-mass members and for which the majority of lower-mass neighbours have not been resolved. For the latter we recommend running the tests of this study a second time, with the desired cluster composition and incompleteness levels, to ascertain the performance of INDICATE and the validity of its index to correctly identify these spatial behaviours. 

We find an inverse correlation between INDICATE’s derived index and cluster binarity in most incompleteness scenarios explored for clusters not associated with their natal nebulosity. Typically, as the binary fraction increased the index values derived for members decreased; this effect was only observed for all resolved member samples, but not in the spatially clustered sub-samples shown in Figure\,\ref{fig_Ipercent},  which is attributable to the resolution capabilities of Gaia. Binary members typically have smaller angular separations than unpaired neighbours so dimmer companions are not resolved, due to contrast separation limitations, but are also more likely to be occluded by the PSFs of high-mass members. This decrease in resolved stars caused a proportion of members,  particularly those with few neighbours,  to experience a non-negligible perceived decrease in their number of neighbours, and thus a drop in their post-observation index value. However, the typical decrease is small in the spatially clustered populations’ overall pre- and post-observation index values in high-binarity clusters, and therefore insufficient to significantly alter any conclusions regarding their spatial behaviour, though they are mildly diluted in most cases. Wide binaries were not explored in this study, but are unlikely to induce a similar change in the pre- to post-observation indices as Gaia is better able to resolve these pairings \citep{2020ApJS..247...66H}. No decrease in pre- and post-observation index values with increasing binarity was found in clusters still associated with their natal nebulosity as in the presence of variable extinction intrinsically bright stars can appear dimmer relative to neighbours thereby  reducing the contrast, so those at smaller angular separations are resolved, and  lessening the impact of their PSFs. 

Several limits were placed on our mock Gaia cluster catalogue, namely the minimum number of members, distances, and extinction ranges. To be included in our study at least 50 stars needed to be resolved in each cluster observation because this is the smallest dataset INDICATE can be run on (below this small number statistics can become significant, \citealt{2019A&A...622A.184B}). Subsequently 2616 out of 6800 Obs-A cluster observations were excluded from further analysis. We chose not to increase the total number of pre-observed cluster members to compensate for this as our clusters were designed to approximate typical young star forming regions; as Gaia is a visual band survey it is realistic to expect that a significant proportion of clusters will not meet the required minimum number of resolved members  to be analysed with INDICATE. This limitation can be overcome in real observational studies if Gaia data is used in combination with a longer wavelength survey (e.g. Spitzer-MIPS, \citealt{2004ApJS..154...25R}; VISTA-VVV, \citealt{2010NewA...15..433M}; VISTA-VHS, \citealt{2013Msngr.154...35M}; UKIDSS, \citealt{2007MNRAS.379.1599L}), but as these are typically not available for all-sky, and we are specifically interested in the effect of the Gaia filter, we did not include simulated data from other surveys in our analysis. Similarly we chose to cap cluster distance at 1\,kpc as dataset incompleteness becomes a significant issue at  greater distances with visual band surveys. 

We explored the effects of uniform and variable visual extinction. For the former a wide range of values (20 mags) were applied to the clusters so that the behaviour of INDICATE's index with uniform extinction could be fully studied, but this ultimately proved unnecessary as even at small distances no cluster with an $A_{\text{v}}>$16\,mag met the required minimum number of members. For the variable visual extinction SPH clouds were used to generate $A_{\text{v}}$ histograms for the clusters that had a  dispersion and shape similar to those observed in the literature for galactic young open star clusters with non-uniform extinction. We refrained from using a realistic physical approximation of a natal cluster environment (such as a gas--cloud simulation), as the aim of this test was to determine how well INDICATE handles patchy incompleteness. Therefore,  we only required plausible extinction maps for the clusters, which was achieved using our $A_{\text{v}}$ histograms method; in addition,  cloud simulations are computationally expensive and have specific initial conditions, so many iterations would be required to ascertain the generic behaviour of INDICATE when applied to an observation of a cluster in nebulosities with various initial conditions (which are also unknown).

With the second and third instalments of the Gaia survey, high accuracy distance, position, and kinetic measurements have become available for an unprecedented number of star clusters. The pay-off for this advancement is not only an inevitable culling of membership lists, but also significant sample incompleteness as Gaia is an optical-wavelength survey. To better understand and characterise spatial behaviours in young clusters it is imperative that the impact of this incompleteness on spatial distribution studies is ascertained, so the correct conclusions are drawn about the properties of clusters. In this work we have shown through extensive statistical testing on a mock Gaia cluster catalogue that the spatial analysis tool INDICATE can be used to robustly study these behaviours in Gaia-observed young star forming regions up to 1\,kpc with an incompleteness level of $83.3\%$ and binarity of $50\%$.

\begin{acknowledgements}

A. Buckner has received funding for this research from the ICYBOB project under the European Research Council H2020-EU.1.1. programme (Grant No. 818940) and the StarFormMapper project under European Union’s Horizon 2020 research \& innovation programme (Grant No. 687528). The authors would like to thank the referee S.Lepine for his constructive and insightful feedback which led to the improvement of the manuscript.

\end{acknowledgements}

\bibliographystyle{aa} 
\bibliography{refs} 

\begin{appendix}
\onecolumn

\section{Reference tables for INDICATE}\label{app_ref}

Section\,\ref{sect_results} described the general trends in change of index values derived by INDICATE for clusters in Obs-A (i.e. as a function of their distance, uniform foreground extinction, and binary fraction). Here we present reference tables specifying the statistics as a function of these three variables for the  general, OB, and  non-OB stellar populations of clusters. Each table lists the cluster distance ($D$), visual extinction ($A_{\text{v}}$), percentage of members not detected by Gaia ($\%\,\text{Mem}_{\text{\,ND}}$);  and with respect to their true values, percentage point change in the number of observed members found to be spatially clustered ($\%\,\text{Mem}_{\text{\,cl}}$), percentage change in the median index value derived for clustered stars ($\%\,\Delta\,\tilde{I}^{\,cl}_5$), percentage change in the median index value derived for all stars ($\%\,\Delta\,\tilde{I}_5$), and percentage change in the maximum index value for a star in the cluster ($\%\,\Delta\,\max{I}_5$).\newline

\clearpage
\twocolumn

\small 
\topcaption{Statistics for the general stellar population of clusters with a binary fraction of zero. \label{table_b0_all}} 
\tablehead{\hline \hline $D$ & $A_{\text{v}}$ & $\%\,\text{Mem}_{\text{\,ND}}$ & $\%\,\text{Mem}_{\text{\,cl}}$ & $\%\,\Delta\,\tilde{I}^{\,cl}_5$ & $\%\,\Delta\,\tilde{I}_5$ & $\%\,\Delta\,\max{I}_5$ \\ \hline} 
\tabletail{\hline}
\begin{supertabular}{c c c c c c c}
100 & 0 & 0.5 & 0.0 & 0.0 & 0.0 & 0.0 \\ 
 100 & 1 & 0.3 & 0.0 & 0.0 & 0.0 & 0.0 \\ 
 100 & 2 & 0.3 & 0.1 & 0.0 & 0.0 & 0.0 \\ 
 100 & 3 & 0.3 & 0.1 & 0.0 & 0.0 & 0.0 \\ 
 100 & 4 & 0.3 & 0.0 & 0.0 & 0.0 & 0.0 \\ 
 100 & 5 & 0.3 & 0.0 & 0.0 & 0.0 & 0.0 \\ 
 100 & 6 & 0.3 & 0.0 & 0.0 & 0.0 & 0.0 \\ 
 100 & 7 & 0.0 & 0.0 & 0.0 & 0.0 & 0.0 \\
\end{supertabular}
\clearpage


\topcaption{Statistics for the OB population of clusters with a binary fraction of zero. \label{table_b0_ob}} 
\tablehead{\hline \hline $D$ & $A_{\text{v}}$ & $\%\,\text{Mem}_{\text{\,ND}}$ & $\%\,\text{Mem}_{\text{\,cl}}$ & $\%\,\Delta\,\tilde{I}^{\,cl}_5$ & $\%\,\Delta\,\tilde{I}_5$ & $\%\,\Delta\,\max{I}_5$ \\ \hline} 
\tabletail{\hline}
\begin{supertabular}{c c c c c c c}
 
 100 & 0 & 11.7 & 0.0 & 0.0 & 0.0 & 0.0 \\ 
 100 & 1 & 9.6 & 0.0 & 0.0 & 0.0 & 0.0 \\ 
 100 & 2 & 8.7 & 0.0 & 0.0 & 0.0 & 0.0 \\ 
 100 & 3 & 8.7 & 0.0 & 0.0 & 0.0 & 0.0 \\ 
 100 & 4 & 4.2 & 0.0 & 0.0 & 0.0 & 0.0 \\ 
 100 & 5 & 4.2 & 0.0 & 0.0 & 0.0 & 0.0 \\ 
 100 & 6 & 4.2 & 0.0 & 0.0 & 0.0 & 0.0 \\ 
 100 & 7 & 0.0 & 0.0 & 0.0 & 0.0 & 0.0 \\
\end{supertabular}
\clearpage


\topcaption{Statistics for the non-OB stellar population of clusters with a binary fraction of zero. \label{table_b0_Nob}}   
\tablehead{\hline \hline $D$ & $A_{\text{v}}$ & $\%\,\text{Mem}_{\text{\,ND}}$ & $\%\,\text{Mem}_{\text{\,cl}}$ & $\%\,\Delta\,\tilde{I}^{\,cl}_5$ & $\%\,\Delta\,\tilde{I}_5$ & $\%\,\Delta\,\max{I}_5$ \\ \hline} 
\tabletail{\hline}

\begin{supertabular}{c c c c c c c}

100 & 0 & 0.0 & 0.0 & 0.0 & 0.0 & 0.0 \\ 
 100 & 1 & 0.0 & 0.0 & 0.0 & 0.0 & 0.0 \\ 
 100 & 2 & 0.0 & 0.0 & 0.0 & 0.0 & 0.0 \\ 
 100 & 3 & 0.0 & 0.0 & 0.0 & 0.0 & 0.0 \\ 
 100 & 4 & 0.0 & 0.0 & 0.0 & 0.0 & 0.0 \\ 
 100 & 5 & 0.0 & 0.0 & 0.0 & 0.0 & 0.0 \\ 
 100 & 6 & 0.0 & 0.0 & 0.0 & 0.0 & 0.0 \\ 
 100 & 7 & 0.0 & 0.0 & 0.0 & 0.0 & 0.0 \\
\end{supertabular}
\clearpage



\topcaption{Statistics for the general stellar population of clusters with a binary fraction of $25\%$. \label{table_b25_all}}    
\tablehead{\hline \hline $D$ & $A_{\text{v}}$ & $\%\,\text{Mem}_{\text{\,ND}}$ & $\%\,\text{Mem}_{\text{\,cl}}$ & $\%\,\Delta\,\tilde{I}^{\,cl}_5$ & $\%\,\Delta\,\tilde{I}_5$ & $\%\,\Delta\,\max{I}_5$ \\ \hline} 
\tabletail{\hline}
\begin{supertabular}{c c c c c c c}

 100 & 0 & 6.7 & -1.5 & 0.0 & 0.0 & -8.1 \\ 
 100 & 1 & 6.7 & -1.6 & 0.0 & 0.0 & -8.1 \\ 
 100 & 2 & 6.7 & -1.6 & 0.0 & 0.0 & -8.1 \\ 
 100 & 3 & 6.7 & -1.6 & 0.0 & 0.0 & -8.1 \\ 
 100 & 4 & 6.7 & -1.6 & 0.0 & 0.0 & -8.1 \\ 
 100 & 5 & 6.7 & -1.6 & 0.0 & 0.0 & -8.1 \\ 
 100 & 6 & 6.7 & -1.6 & 0.0 & 0.0 & -8.1 \\ 
 100 & 7 & 6.3 & -1.6 & -3.1 & 0.0 & -8.1 \\
\end{supertabular}
\clearpage


\topcaption{Statistics for the OB population of clusters with a binary fraction of $25\%$. \label{table_b25_ob}}
\tablehead{\hline \hline $D$ & $A_{\text{v}}$ & $\%\,\text{Mem}_{\text{\,ND}}$ & $\%\,\text{Mem}_{\text{\,cl}}$ & $\%\,\Delta\,\tilde{I}^{\,cl}_5$ & $\%\,\Delta\,\tilde{I}_5$ & $\%\,\Delta\,\max{I}_5$ \\ \hline} 
\tabletail{\hline}
\begin{supertabular}{c c c c c c c}

100 & 0 & 19.8 & -4.9 & 3.7 & -8.8 & -4.4 \\ 
 100 & 1 & 14.1 & -6.7 & 0.0 & -9.4 & -4.4 \\ 
 100 & 2 & 9.2 & -7.7 & 3.7 & -9.4 & -4.4 \\ 
 100 & 3 & 9.2 & -7.7 & 3.7 & -9.4 & -4.4 \\ 
 100 & 4 & 8.7 & -7.7 & 3.7 & -9.4 & -4.4 \\ 
 100 & 5 & 8.7 & -7.7 & 3.7 & -9.4 & -4.4 \\ 
 100 & 6 & 8.7 & -7.7 & 3.7 & -9.4 & -4.4 \\ 
 100 & 7 & 0.0 & -8.0 & 3.7 & -5.3 & -4.4 \\
\end{supertabular}
\clearpage


\topcaption{Statistics for the non-OB stellar population of clusters with a binary fraction of $25\%$. \label{table_b25_Nob}}   
\tablehead{\hline \hline $D$ & $A_{\text{v}}$ & $\%\,\text{Mem}_{\text{\,ND}}$ & $\%\,\text{Mem}_{\text{\,cl}}$ & $\%\,\Delta\,\tilde{I}^{\,cl}_5$ & $\%\,\Delta\,\tilde{I}_5$ & $\%\,\Delta\,\max{I}_5$ \\ \hline} 
\tabletail{\hline}
\begin{supertabular}{c c c c c c c}
  
100 & 0 & 6.1 & -1.8 & -4.8 & 0.0 & -8.1 \\ 
 100 & 1 & 6.2 & -1.8 & -4.8 & 0.0 & -8.1 \\ 
 100 & 2 & 6.2 & -1.8 & -4.8 & 0.0 & -8.1 \\ 
 100 & 3 & 6.2 & -1.8 & -4.8 & 0.0 & -8.1 \\ 
 100 & 4 & 6.2 & -1.8 & -4.8 & 0.0 & -8.1 \\ 
 100 & 5 & 6.2 & -1.8 & -4.8 & 0.0 & -8.1 \\ 
 100 & 6 & 6.2 & -1.8 & -4.8 & 0.0 & -8.1 \\ 
 100 & 7 & 6.2 & -1.5 & -4.8 & 0.0 & -8.1 \\
\end{supertabular}
\clearpage



\topcaption{Statistics for the general stellar population of clusters with a binary fraction of $50\%$. \label{table_b50_all}}    
\tablehead{\hline \hline $D$ & $A_{\text{v}}$ & $\%\,\text{Mem}_{\text{\,ND}}$ & $\%\,\text{Mem}_{\text{\,cl}}$ & $\%\,\Delta\,\tilde{I}^{\,cl}_5$ & $\%\,\Delta\,\tilde{I}_5$ & $\%\,\Delta\,\max{I}_5$ \\ \hline} 
\tabletail{\hline}
\begin{supertabular}{c c c c c c c}
 
 100 & 0 & 13.7 & -1.4 & -2.6 & 0.0 & -2.8 \\ 
 100 & 1 & 13.7 & -1.4 & -2.6 & 0.0 & -2.8 \\ 
 100 & 2 & 13.5 & -1.4 & -2.6 & 0.0 & -2.8 \\ 
 100 & 3 & 13.5 & -1.9 & -2.6 & 0.0 & -2.8 \\ 
 100 & 4 & 13.5 & -1.9 & -2.6 & 0.0 & -2.8 \\ 
 100 & 5 & 13.5 & -2.4 & -2.6 & 0.0 & -2.8 \\ 
 100 & 6 & 13.5 & -2.4 & -2.6 & 0.0 & -2.8 \\ 
 100 & 7 & 13.3 & -2.7 & -2.6 & 0.0 & -1.3 \\
\end{supertabular}
\clearpage


\topcaption{Statistics for the OB population of clusters with a binary fraction of $50\%$. \label{table_b50_ob}}  
\tablehead{\hline \hline $D$ & $A_{\text{v}}$ & $\%\,\text{Mem}_{\text{\,ND}}$ & $\%\,\text{Mem}_{\text{\,cl}}$ & $\%\,\Delta\,\tilde{I}^{\,cl}_5$ & $\%\,\Delta\,\tilde{I}_5$ & $\%\,\Delta\,\max{I}_5$ \\ \hline} 
\tabletail{\hline}
\begin{supertabular}{c c c c c c c}
 100 & 0 & 29.7 & -13.6 & 5.1 & -19.1 & -3.7 \\ 
 100 & 1 & 22.6 & -10.8 & -2.6 & -19.1 & -3.7 \\ 
 100 & 2 & 21.1 & -10.8 & -2.6 & -14.4 & -3.7 \\ 
 100 & 3 & 21.1 & -10.8 & -2.6 & -14.4 & -3.7 \\ 
 100 & 4 & 21.1 & -10.8 & -2.6 & -12.3 & -3.7 \\ 
 100 & 5 & 21.1 & -10.8 & -2.6 & -12.3 & -3.7 \\ 
 100 & 6 & 21.1 & -10.8 & -2.6 & -12.3 & -3.7 \\ 
 100 & 7 & 14.9 & -10.3 & 7.7 & -16.6 & 4.1 \\
\end{supertabular}
\clearpage


\topcaption{Statistics for the non-OB stellar population of clusters with a binary fraction of $50\%$. \label{table_b50_Nob}}    
\tablehead{\hline \hline $D$ & $A_{\text{v}}$ & $\%\,\text{Mem}_{\text{\,ND}}$ & $\%\,\text{Mem}_{\text{\,cl}}$ & $\%\,\Delta\,\tilde{I}^{\,cl}_5$ & $\%\,\Delta\,\tilde{I}_5$ & $\%\,\Delta\,\max{I}_5$ \\ \hline} 
\tabletail{\hline}
\begin{supertabular}{c c c c c c c}

100 & 0 & 12.9 & -2.3 & -1.5 & 0.0 & -2.8 \\ 
 100 & 1 & 13.1 & -2.3 & -1.5 & 0.0 & -2.8 \\ 
 100 & 2 & 12.9 & -2.0 & -2.6 & 0.0 & -2.8 \\ 
 100 & 3 & 12.9 & -2.0 & -2.6 & 0.0 & -2.8 \\ 
 100 & 4 & 13.1 & -2.0 & -2.6 & 0.0 & -2.8 \\ 
 100 & 5 & 13.1 & -2.7 & -2.6 & 0.0 & -2.8 \\ 
 100 & 6 & 13.1 & -2.7 & -2.6 & 0.0 & -2.8 \\ 
 100 & 7 & 13.1 & -2.9 & -2.6 & 0.0 & -1.3 \\
\end{supertabular}
\clearpage


\end{appendix}

\end{document}